\documentclass[11pt]{article}
\usepackage{amsmath}
\usepackage{amssymb}
\usepackage[mathscr]{eucal}
\usepackage[usenames]{color}
\usepackage{url}
\usepackage[utf8]{inputenc}
\usepackage{graphicx}
\usepackage{epstopdf}
\usepackage{setspace}
\usepackage{color}
\usepackage{subfigure}
\usepackage{tikz}
\usetikzlibrary{shapes.geometric, arrows}
\tikzstyle{startstop} = [rectangle, rounded corners, minimum width=3cm, minimum height=1cm,text centered, draw=black, fill=red!30]
\usepackage{authblk}

\usepackage[square,numbers]{natbib}

\definecolor{webgreen}{rgb}{0,0.5,0}

\numberwithin{equation}{section}
\numberwithin{thm}{section}
\numberwithin{lemma}{section}
\numberwithin{prop}{section}
\numberwithin{cor}{section}
\numberwithin{rmk}{section}
\numberwithin{defn}{section}

\definecolor{darkolivegreen}{rgb}{0.333333, 0.419608, 0.1843140}

\begin{document}
\pagenumbering{arabic}

\title{\Large A New Susceptible-Infectious (SI) Model  With Endemic Equilibrium}

\author[1,2]{Ayse Peker Dobie\thanks{pdobie@itu.edu.tr- Corresponding Author}}
\author[1]{Semra Ahmetolan \thanks{ahmetola@itu.edu.tr}}
\author[2]{Ayse Humeyra Bilge \thanks{ayse.bilge@khas.edu.tr}}
\author[1]{Ali Demirci \thanks{demircial@itu.edu.tr}}
\affil[1]{\small Department of Mathematics, Faculty of Science and Letters, Istanbul Technical University, Istanbul, Turkey}
\affil[2]{TDepartment of Industrial Engineering, Faculty of Engineering and Natural Sciences,  Kadir Has  University, Istanbul, Turkey }

\renewcommand\Authands{ and }

\date{\today}
\clearpage
\maketitle
\thispagestyle{empty}
\begin{abstract}
The focus of this article is on the  dynamics of a new susceptible-infected model which consists of a susceptible group ($S$) and  two different infectious groups ($I_1$ and $I_2$). Once infected, an individual becomes a member of one of these infectious groups which have different clinical forms of  infection. In addition, during the progress of the illness, an infected individual in group $I_1$ may pass to the infectious group $I_2$ which has a higher mortality rate.  In this study, positiveness  of the solutions for the model is proved. Stability analysis of  species extinction, $I_1$-free equilibrium and endemic equilibrium as well as disease-free equilibrium is studied. Relation between the basic reproduction number of the disease and the basic reproduction number of each infectious stage is examined. The model is investigated  under a specific condition and  its exact solution  is  obtained.

\end{abstract}

\section{Introduction}
One major contribution of mathematics to epidemiology is the compartmental model introduced by Kermack and McKendrick  in 1927  \cite{ker}. Since then, significant progress has been achieved and numerous mathematical models have been developed to study various  diseases \cite{cia, whi}. Major outbreaks such as SARS epidemic in 2002 \cite{cho, gum, bau}, H5N1 influenza epidemic in 2005 \cite{rao, yan, che}, H1N1 influenza pandemic in 2009 \cite{pek, dem, cob}, Ebola in 2014 \cite{chr, webb, mam}, and nowadays Covid-19 pandemic \cite{ahm, kuc, liu} maintain the high interest in mathematical modelling and analysis of infectious diseases.

The  Susceptible-Infected-Removed  ($SIR$),  the  Susceptible-Infected-Susceptible  ($SIS$)  and  the Susceptible-Infected ($SI$) systems are the three fundamental epidemic models \cite{and, mur}. In the $SIR$ model, infected individuals gain permanent immunity after recovering from the disease whereas in the $SIS$ model, they return to the susceptible group. On the other hand,  in  the $SI$ model, infected individuals are lifetime infectious. Thus, the $SI$ model consists of two different compartments; the susceptible group whose members are not yet infected by the pathogen and the infected group whose members are infected by the pathogen.  Various infectious diseases such as AIDS caused by human immune deficiency virus (HIV) and Feline Infectious Peritonities (FIP) in cats caused by Feline Corona virus (FCoV) have more than one infectious stage. In this article,  a new susceptible-infected model ($SI_1I_2$) which has two infectious stages is considered. 

The $SI_1I_2$ model consists of three groups: susceptible population $S$, $I_1$-infected group and $I_2$-infected group. Each individual in the population is in one of these groups since individuals never leave the infectious group once infected and each healthy newborn has no immunity. Once susceptible individuals are infected, they develop the disease in two different clinical forms, $I_1$ and $I_2$. Each infected group  contributes to its own infected group by transmitting the disease to  susceptible ones. In addition, $I_1$-infected individuals may develop the infection to a further stage and may become a member of the group $I_2$ which has a higher mortality rate. Healthy newborns are only seen in the susceptible population. The  newborns from   $I_1$ and $I_2$-infected mothers  are members of the same infected group as their mothers. 

The article consists of five sections. In Section 2, the mathematical model of the disease is described. Theoretical results of the model about the basic reproduction number, disease free equilibrium, endemic equilibrium, extinction, and the phase portraits  are presented in Section 3.  The relation between the basic reproduction number of each form of the infectious disease  is also given in this section. A model of the disease   under specific conditions  is presented in Section 4. Exact solutions of this reduced model are also obtained in this section.  Concluding remarks and the discussion of the results are given in Section 5.

\section{$SI_1I_2$ Model}

The susceptible-infected  model  considered in this work is described by the  following system of nonlinear ordinary differential equations
\begin{eqnarray}
\begin{aligned}
S'&=-\beta_1 SI_1-\beta_2 SI_2+f_0 S\\
I_1'&=\beta_1 SI_1-\theta I_1+f_1 I_1\\
I_2'&=\beta_2 SI_2+\theta I_1+f_2 I_2
\end{aligned}
\end{eqnarray}
where $f_0=\mu_0-\delta_0$, $f_1=\mu_1-\delta_1$ and $f_2=\mu_2-\delta_2$, and $\mu_i$ and $\delta_i$ are the  birth and death rates of each group. In this model denoted by  $SI_1I_2$ (figure 1), $S$ represents the susceptible population whereas $I_1$ and $I_2$ are the two different groups of the  population infected by the same virus but which develop the disease in different clinical forms.  Each infected population comes into contact with the susceptible population with different contact rates, $\beta_i$.  The parameter $\theta$ is the rate of the individuals in the group $I_1$ who become a member of the group $I_2$. The parameters $\beta_i$, $\mu_i$, $\delta_i$  and $\theta$   are positive and the nonconstant total population size $N$ is equal to $S+I_1+I_2$. Note that it will be necessary to choose 
\begin{eqnarray}
\begin{aligned}
\theta-f_1>0\nonumber\\  f_2<0,\nonumber\\
\end{aligned}
\end{eqnarray}
otherwise $I_1$ and $I_2$ will be unbounded.

\begin{figure}[ht]
\centering
\includegraphics[scale=0.7]{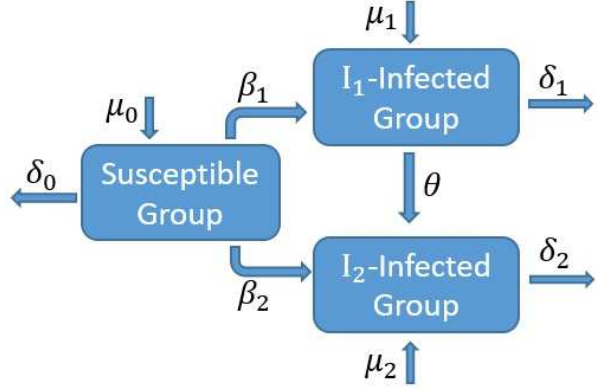}
\caption{Diagram of the $SI_1I_2$ model. Parameters $\beta_i$, $\mu_i$ and $\delta_i$ are the transmission, birth and death rates of the group $I_i$, respectively whereas $\mu_0$ and $\delta_0$ are the birth and death rates of the group $S$. The parameter $\theta$ is the rate of the individuals in the group $I_1$ who become a member of the group $I_2$.}
\end{figure}

The  $SI_1I_2$ model is based on the following assumptions:

$\bullet$ Each individual in the population is in one of these three classes, $S$, $I_1$ or $I_2$.

$\bullet$ Infected individuals remain infective for the rest of their lives.

$\bullet$ Newborns belong to the compartment of their mothers.

$\bullet$ It is  assumed  that each infected group $I_i$ contributes to its own infected group by transmitting the disease to  susceptible individuals. This means that if   individuals from the susceptible group are infected by a member from the group $I_i$ they become a member of that infected group.

$\bullet$ There is a flow from the group $I_1$ to the group $I_2$ since it is assumed that the members in the group $I_1$ may  develop the disease form of the group $I_2$. However, there is no flow from the group $I_2$ to the group $I_1$.

$\bullet$ The distinction between disease related and natural deaths is taken into account by different mortality rates of each group.

The $SI_1I_2$ model admits solutions that are monotone, oscillatory or with decay in oscillatory as shown in Figure 2.

\begin{figure}[ht]
\centering
\includegraphics[scale=0.37]{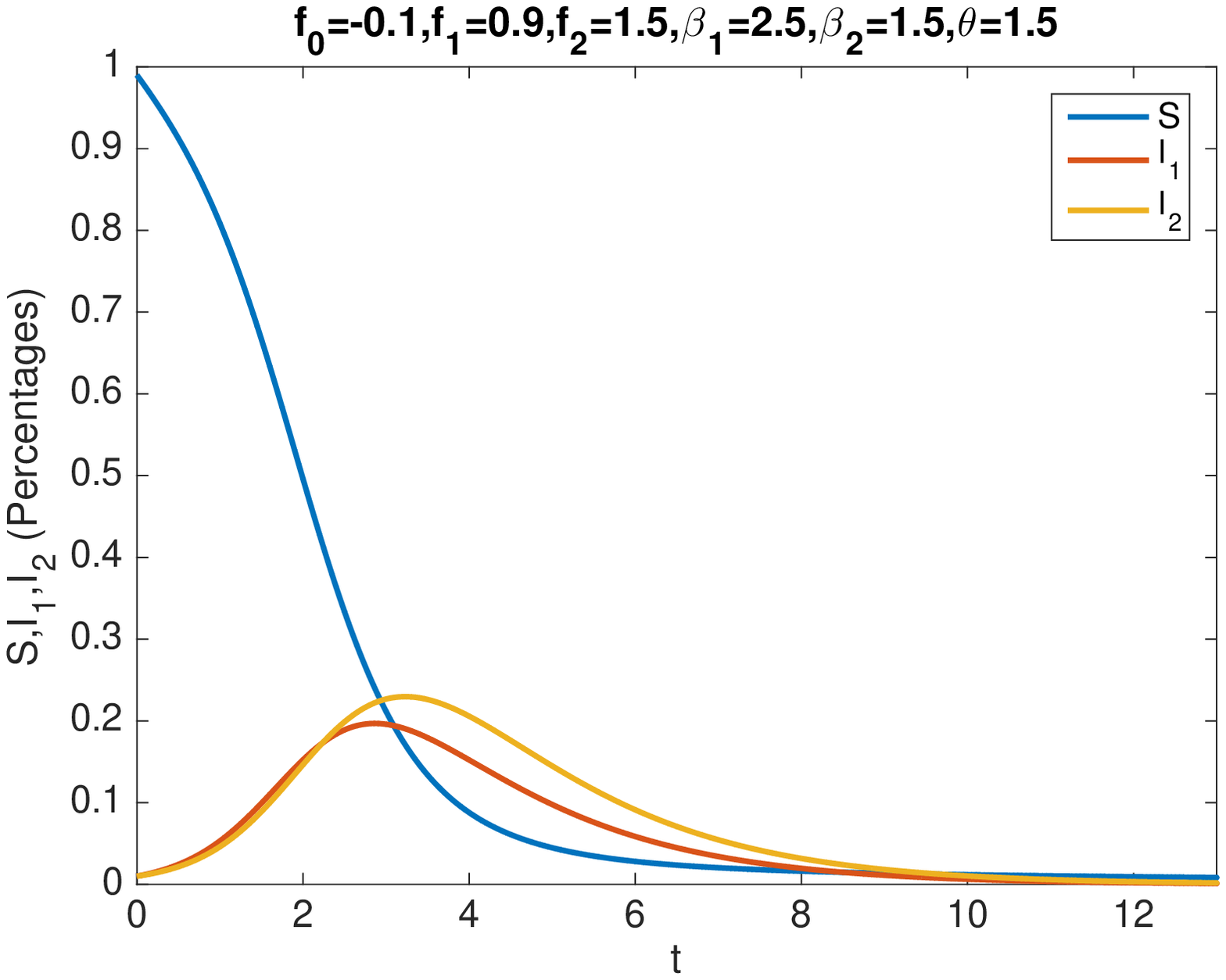}
\includegraphics[scale=0.37]{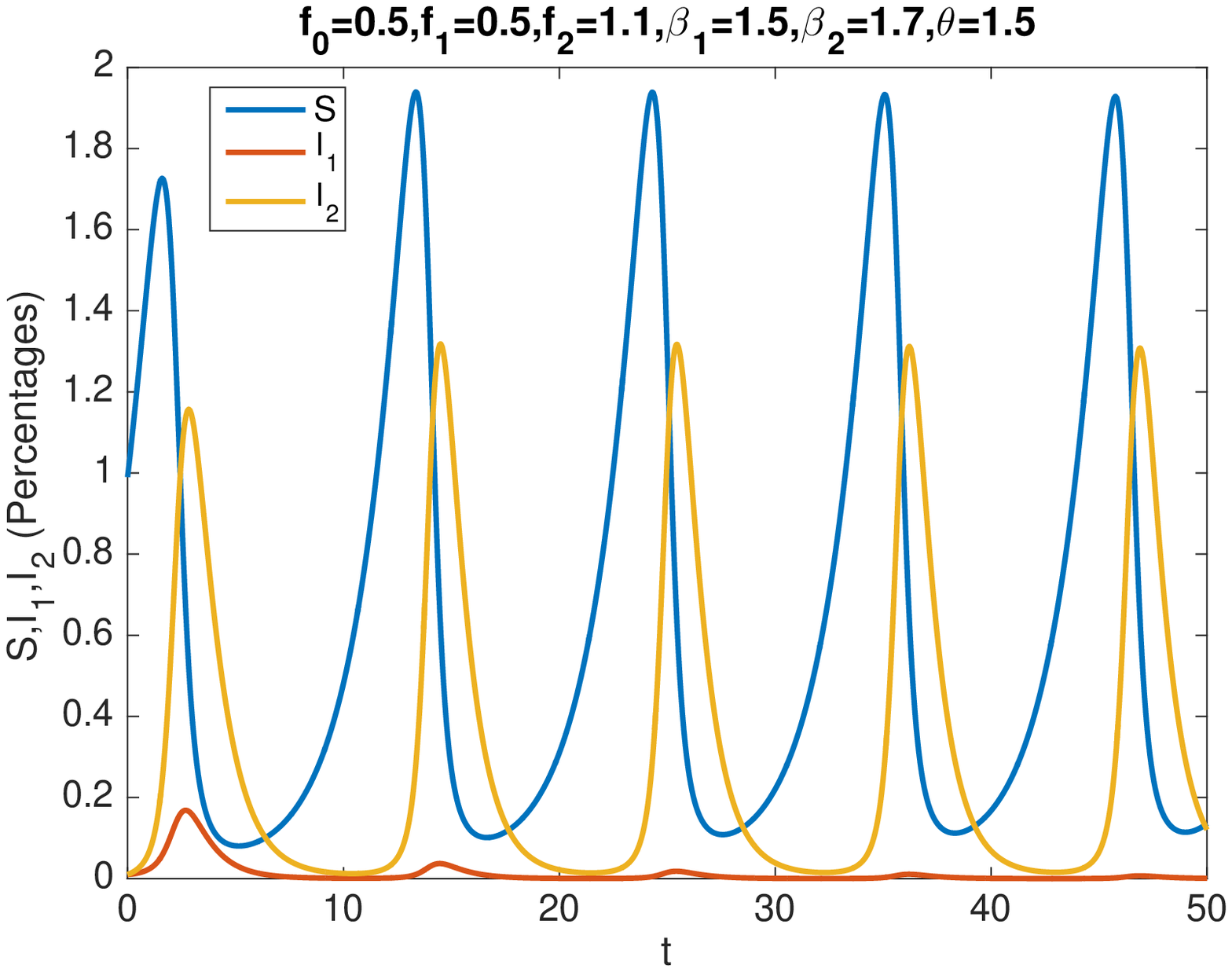}
\includegraphics[scale=0.35]{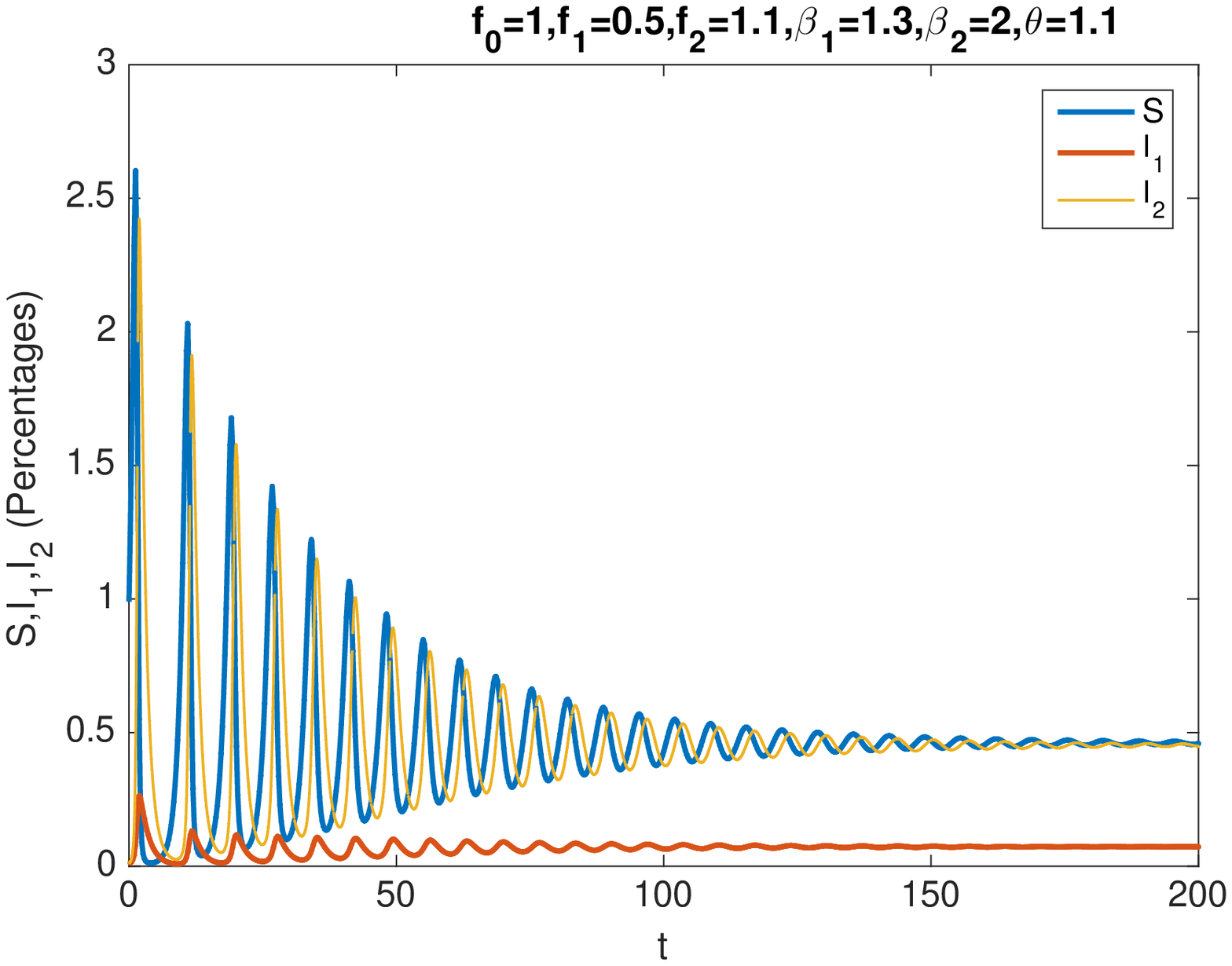}
\caption{Solutions of   different types  for the $SI_1I_2$ model.  }
\end{figure}

\section{Theoretical Results}
In this section, positiveness of the model is proved and the basic reproduction number is found.  The disease free equilibrium at which the population remains in the absence of disease, and the endemic equilibrium are examined.

\subsection{Positiveness}

\textbf{Proposition 1:} The susceptible group $S$ and the infected groups $I_1$ and $I_2$ stay positive if the initial conditions are chosen to be positive.

\vskip.2cm
\textbf{Proof:} The first and the second  equations in (2.1) can be expressed as follows
\begin{eqnarray}
{S'}/{S}&=&-\beta_1I_1-\beta_2I_2+f_0\nonumber\\
{I_1'}/{I_1}&=&\beta_1S-\theta+f_1.\nonumber
\end{eqnarray}

Integration of the  equations above gives
\begin{eqnarray}
S (t)&=& S_0 \,\exp {\bigg(\displaystyle\int_{0}^{t} (-\beta_1I_1-\beta_2I_2+f_0)\, d\tau}\bigg)\nonumber\\
I_1 (t)&=& I_{1,0} \,\exp {\bigg(\displaystyle\int_{0}^{t} (\beta_1S-\theta+f_1)\, d\tau}\bigg)
\end{eqnarray}
where $S_0$ and $I_{1,0}$ are the initial conditions for the susceptible population $S$ and the infected group $I_1$, respectively. As a consequence of the equations in (3.1), $S(t)$ and $I_1(t)$ remain positive if the initial conditions are chosen to be positive.

If  the last equation in (2.1) is integrated, one finds 
\begin{eqnarray}
I_2(t)= \exp {\bigg(\displaystyle\int_{0}^{t} (\beta_2S+f_2)\, d\tau\bigg)}\bigg[ I_{2,0} + \displaystyle\int_{0}^{t} I_1(\tau) \exp {\bigg(-\displaystyle\int_{0}^{\tau} (\beta_2S+f_2)\, d\sigma\bigg)}d\tau\bigg].
\end{eqnarray}
The equation (3.2) yields that $I_2(t)$  remains positive if the initial condition $I_{2,0}$ is chosen to be positive.







\subsection{Basic Reproduction Number}
One of the key parameters in mathematical modelling of transmissible diseases is the basic reproduction number ($R_0$)  which is defined as  the number of new infectious individuals produced by a typical infective individual in a fully susceptible population at a disease-free equilibrium. For models with more than one infected compartments, the computation of  $R_0$ is based on the construction of the next generation matrix whose $(i,j)$ entry is the expected number of new infections in the compartment $i$ produced by the infected individual in the compartment $j$. For mutistrain models, the maximum of the eigenvalues of this  matrix is the basic reproduction number  \cite{Van, Die}. 

For the $SI_1I_2$ model,  let $I^*_{1,0}=I^*_{2,0}=0$ and $S^*_0$ be the disease free equilibrium. If the system of ordinary differential equations in (2.1) is linearized about the disease free equilibrium one obtains the following linearized infection subsystem
\begin{eqnarray}
\begin{aligned}
I_1'&=\beta_1 S^*_0 I_1-\theta I_1+f_1I_1\\
I_2'&=\beta_2 S^*_0 I_2+\theta I_1+f_2 I_2.
\end{aligned}
\end{eqnarray}

The  matrices   $F=\left(
                \begin{array}{cc}
                  \beta_1 S^*_0  & 0 \\
                  0 & \beta_2  S^*_0 \\
                \end{array}\right)$ and  $V=\left(
                \begin{array}{cc}
                  -\theta+f_1  & 0 \\
                  \theta & f_2 \\
                \end{array}
              \right)$
are defined  by using (3.3) such that $K'=(F+V)K$ where $K=\left(
                \begin{array}{c}
                  I_1 \\ 
         I_2 \\
         \end{array}
              \right)$. Note that the $i$th component of the matrix $FK$ is the rate of appearance of new infected individuals in the group $I_i$ whereas the $i$th component of the matrix $VK$ is the rate of transfer of the individuals into and out of  the group $I_i$. Then the next generation matrix is used to derive the basic reproduction number
\begin{eqnarray}
\begin{aligned}
L=-F\cdot V^{-1}=\left(
                \begin{array}{cc}
                  \displaystyle\frac{\beta_1 S^*_0 }{\theta-f_1} & 0 \\ \\
                  -\displaystyle\frac{\beta_2 S^*_0 \theta}{f_2(\theta-f_1)} & -\displaystyle\frac {\beta_2 S^*_0 }{f_2} \\
                \end{array}\right)
\end{aligned}
\end{eqnarray}
where $L_{ij}$ is the number of secondary infections caused in the compartment $i$ by an infected individual in the compartment $j$ \cite{Van, Die}. Thus the eigenvalues of the next generation matrix are found to be
\begin{eqnarray}
\begin{aligned}
\lambda_1=\displaystyle\frac{\beta_1 S^*_0 }{\theta-f_1}\,,\,\lambda_2=-\displaystyle\frac {\beta_2 S^*_0 }{f_2}.
\end{aligned}
\end{eqnarray}
Then the basic reproduction number which is the number of secondary cases produced by a single infective individual introduced into a population is the largest eigenvalue of the matrix $L$; that is,
\begin{eqnarray}
\begin{aligned}
R_0=\max\{\lambda_1,\lambda_2\}.
\end{aligned}
\end{eqnarray}

Here, $\lambda_1$ and $\lambda_2$ are the basic reproduction numbers for each strain corresponding to the groups  $I_1$ and $I_2$, respectively. For the rest of the article, in order to distinguish the basic reproduction numbers for each strain, $\lambda_1$ and $\lambda_2$ will be denoted by $R_1$ and $R_2$, respectively.

\subsection{Disease Free Equilibrium}
\textbf{Proposition 2:} The disease free equilibrium  is locally asymptotically stable if the following conditions are satisfied
\begin{eqnarray}
\begin{aligned}
f_0&<0\nonumber\\
\beta_1 S^*_0&<\theta-f_1\nonumber\\
\beta_2 S^*_0&<-f_2.\nonumber\\
\end{aligned}
\end{eqnarray}

\textbf{Proof:} To determine the stability of the disease free equilibrium, the Jacobian matrix of the nonlinear system (2.1)  is considered

\begin{eqnarray}
\begin{aligned}
J=\left(
                \begin{array}{ccc}
                  -\beta_1I^*_1-\beta_2I^*_2+f_0  & -\beta_1S^* & -\beta_2S^* \\
                  \beta_1I^*_1  & \beta_1S^*-\theta+f_1 & 0 \\
                  \beta_2I^*_2  & \theta & \beta_2S^*+f_2 \\
                \end{array}\right).
\end{aligned}
\end{eqnarray}

The eigenvalues of the Jacobian matrix at the disease free equilibrium  $(S^*,I^*_1,I^*_2)=(S^*_0,0,0)$ are $\lambda^*_1=f_0$, $\lambda^*_2=\beta_1 S^*_0-\theta+f_1$ and $\lambda^*_3=\beta_2 S^*_0+f_2$. Since all the eigenvalues are negative, the disease free equilibrium is locally asymptotically stable. 

To express the conditions  for asymptotic stability at the disease free equilibrium in terms of the basic production number, one can express  $\lambda^*_2 =(\theta-f_1)(R_1-1)$ and  $\lambda^*_3=-f_2(R_2-1)$. There are two possible cases:

$\bullet$ Case 1: If $R_2<R_1$, then $R_0=R_1$ by (3.6). Thus, if $R_2<R_0=R_1<1$, it means that $\lambda^*_2$ and hence $\lambda^*_3$ are negative.

$\bullet$  Case 2: Similarly, if $R_1<R_2$, then $R_0=R_2$. Thus, if $R_1<R_0=R_2<1$, it means that $\lambda^*_3$ and hence $\lambda^*_2$ are negative.

Since $f_0$ is negative, the analysis of these two cases shows that  the disease free equilibrium   is unstable if $R_0>1$; that is, the invasion of the disease is always possible. The disease free equilibrium is locally asymptotically stable if $R_0<1$. In other words, if $R_1<1$ and $R_2<1$, the disease can not invade the population. This means that  if $R_0<1$, then solutions with initial values close to the disease free equilibrium remain close to this equilibrium and approach to this equilibrium as $t$ approaches infinity.

\subsection{Endemic Equilibrium}

\textbf{Proposition 3:} The following statements hold for  the system defined by the equations in (2.1).

 (i) Species extinction  equilibrium at the point $(S^*,I^*_1,I^*_2)=(0,0,0)$ is locally asymptotically stable if $f_0<0$.
   
  (ii) $I_1$-free equilibrium  at the point $(S^*,I^*_1,I^*_2)=\bigg(-\displaystyle \frac{f_2}{\beta_2},0,\displaystyle \frac{f_0}{\beta_2}\bigg)$ is  stable if $f_0>0$ and  $-\beta_1 f_2 < \beta_2(\theta-f_1)$.
  
  (iii) Endemic equilibrium  at the point $(S^*,I^*_1,I^*_2)=\bigg(\displaystyle \frac{\theta-f_1}{\beta_1},
\displaystyle \frac{f_0}{\beta_1}\bigg(1-\displaystyle \frac{\beta_2\theta}{\beta_2f_1-\beta_1 f_2}\bigg),\displaystyle \frac{\theta f_0}{\beta_2f_1-\beta_1 f_2}\bigg)$ is  locally asymptotically stable if $f_0>0$, $-\beta_1 f_2 > \beta_2(\theta-f_1)$ and $\beta_2>\beta_1$. This equilibrium  is stable if  $f_0>0$, $-\beta_1 f_2 > \beta_2(\theta-f_1)$ and $\beta_2=\beta_1$.

\textbf{Proof:}
To find the equilibrium points, the right hand side of each  equation in (2.1) is set as 0. If $(S^*,I^*_1,I^*_2)$ denote the ordered triple, then the following three critical points are found
\begin{eqnarray}
\begin{aligned}
A(0,0,0),\,\,B\bigg(-\displaystyle \frac{f_2}{\beta_2},0,\displaystyle \frac{f_0}{\beta_2}\bigg),\,\,C\bigg(\displaystyle \frac{\theta-f_1}{\beta_1},
\displaystyle \frac{f_0}{\beta_1}\bigg(1-\displaystyle \frac{\beta_2\theta}{\beta_2f_1-\beta_1 f_2}\bigg),\displaystyle \frac{\theta f_0}{\beta_2f_1-\beta_1 f_2}\bigg).
\end{aligned}
\end{eqnarray}

\vskip.3cm
(i)
  The point $A$ represents the extinction of population; that is, the final population size $N$ is zero. If the Jacobian matrix in (3.7) is evaluated at the point $A$, one finds the matrix
$$\begin{aligned}
\left(
                \begin{array}{ccc}
                  f_0  & 0 & 0 \\
                  0  & -\theta+f_1 & 0 \\
                  0  & \theta & f_2 \\
                \end{array}\right)
\end{aligned}$$
whose eigenvalues  are $\lambda^*_1=f_0$, $\lambda^*_2=f_1-\theta$ and $\lambda^*_3=f_2$. Note that $\lambda^*_2=- {\beta_1 S_0}/{R_1}$ and $\lambda^*_3=- {\beta_2 S_0}/{R_2}$.

Since all the eigenvalues are real, and $\lambda^*_2$ as well as $\lambda^*_3$ are already negative, the stability at this equilibrium  point  depends entirely on the sign of the coefficient $f_0$.  Therefore, the equilibrium at $A$ which will lead to the species extinction is locally asymptotically stable if $f_0$  is negative; that is, the death rate is greater than the birth rate in the susceptible population. This case is illustrated on the top left panel in Figure 2.



\vskip.3cm
(ii)  The point $B$ gives $I_1$-free  equilibrium. Certainly, this equilibrium exists only if the components $S^*$ and $I^*_2$ of $B$ are strictly positive; that is, $f_0>0$. In other words,  the births must be greater than the deaths in the susceptible population for the existence of $I_1$-free  equilibrium.

If the Jacobian matrix in (3.7) is evaluated at the point $B$, one finds the matrix
$$\begin{aligned}
\left(
                \begin{array}{ccc}
                  0  & {\beta_1 f_2}/{\beta_2} & f_2 \\
                  0  & -({\beta_1 f_2}/{\beta_2})-\theta+f_1 & 0 \\
                  f_0  & \theta & 0 \\
                \end{array}\right)
\end{aligned}$$
whose eigenvalues  are
  $\lambda^*_1=-\displaystyle \frac {\beta_1 f_2}{\beta_2}-\theta+f_1$ and $\lambda^*_{2,3}=\mp \sqrt{-f_0f_2}\,i$. 
  
If two of the eigenvalues are pure imaginary complex conjugate numbers, then $\lambda^*_1$ must be negative in order that the linear system is stable; that is, $-\beta_1 f_2 < \beta_2(\theta-f_1)$. 

To express the conditions for stability at the $I_1$-free equilibrium in terms of the basic production number, one can express $\lambda^*_1=\beta_1 S_0\bigg(\displaystyle \frac {1}{R_2}-\displaystyle \frac {1}{R_1}\bigg)$ and $\lambda^*_{2,3}=\mp  \sqrt{\displaystyle \frac  {f_0\beta_2 S_0}{R_2}} \,i$. Thus, if $R_0=R_2$, the linear system is stable at the point $B$.

Stability at the $I_1$-free equilibrium is illustrated on the top right panel in Figure 2. Suitable parameters are chosen  for such an equilibrium at which group $I_1$ has gone extinct with time, whereas  $S$ and $I_2$ are stable over time. Therefore, as  can be observed in this figure, the groups $S$ and $I_2$ exhibit periodic behaviour.

\begin{figure}[ht]
\centering
\includegraphics[scale=0.5]{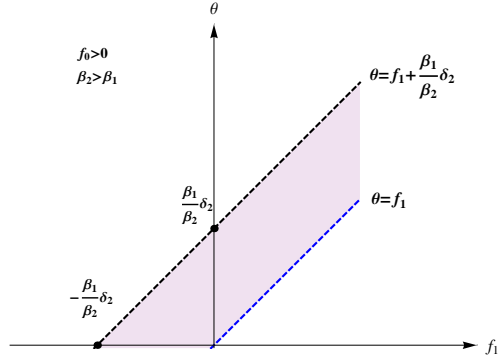}
\caption{The shaded region gives  $(f_1,\theta)$ pairs for which the endemic equilibrium at the point $C$ is always locally asymptotically stable if $f_0>0$ and $\beta_2>\beta_1$.}
\end{figure}

\vskip.3cm
(iii)  The point $C$ represents endemic equilibrium, the existence of which requires  $f_0>0$ and $-\beta_1 f_2 > \beta_2(\theta-f_1)$.

The determinant of the matrix $J-\lambda I$ at the point $C$ is
\begin{eqnarray}
\begin{aligned}
\Delta=-\lambda^{*^3}+(\beta_2S^*+ f_2)\lambda^{*^2}+S^*(-\beta^2_1I^*_1-\beta^2_2I^*_2)\lambda^*+\beta_1I^*_1S^*(\beta_1\beta_2S^*-\beta_1S
-\beta_2\theta).
\end{aligned}
\end{eqnarray}

\begin{figure}[ht]
\centering
\includegraphics[scale=0.37]{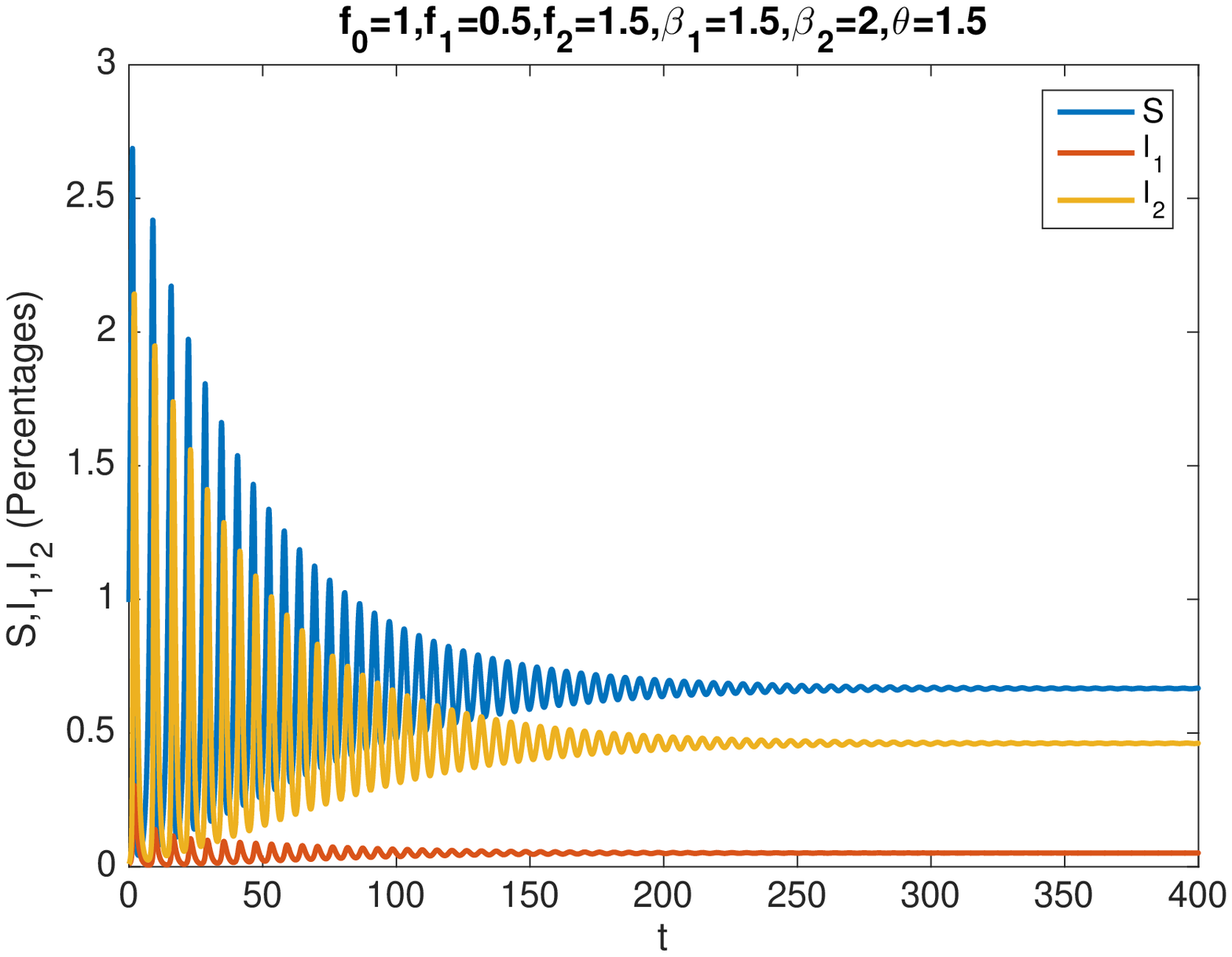}
\includegraphics[scale=0.36]{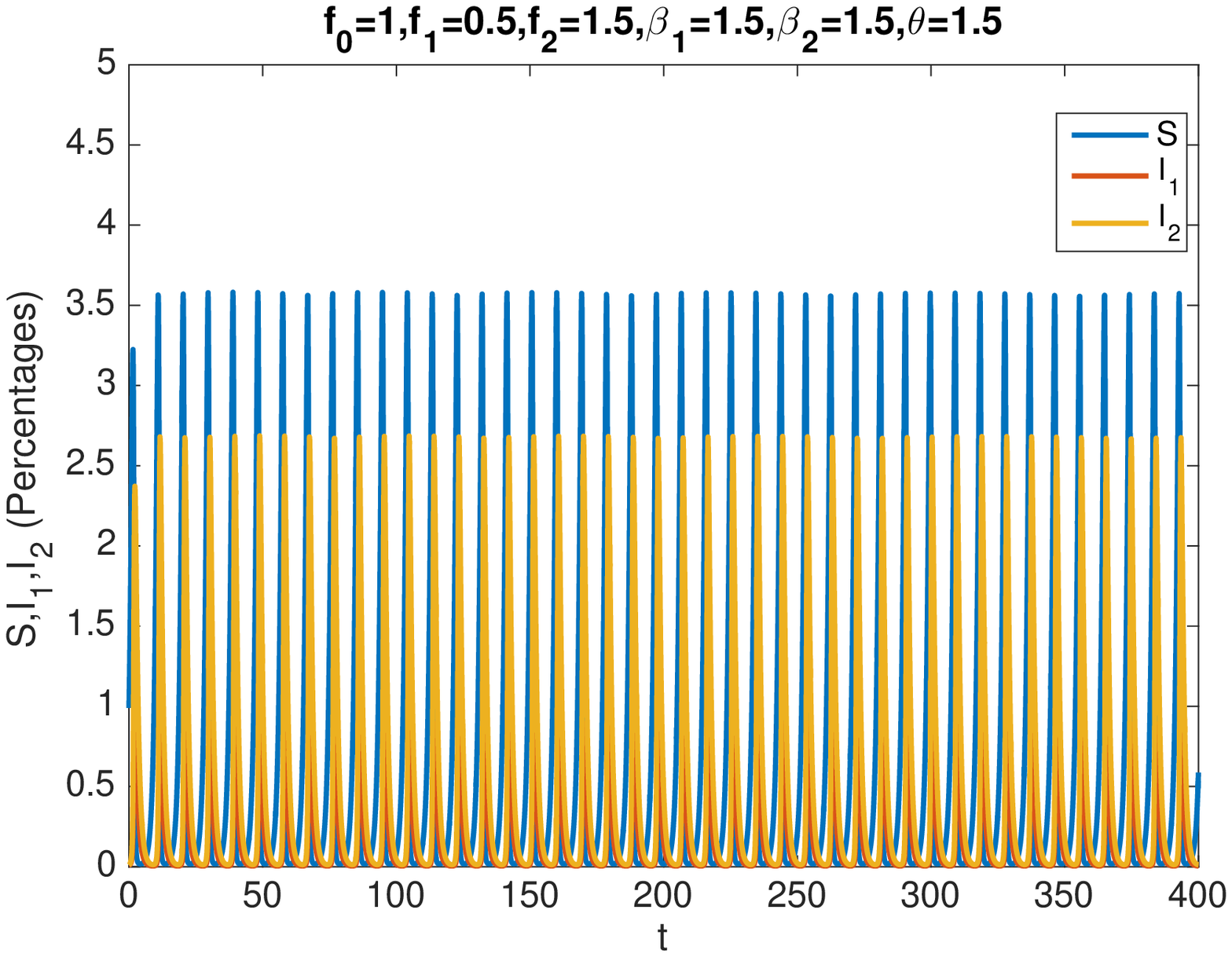}
\caption{Asymptotic stability and stability at the endemic equilibrium   on the left and right panels, respectively.}
\end{figure}

Rather than finding the roots of this cubic equation in (3.9), the Routh-Hurwitz criterion is used to determine the character of the roots. This criterion states that all the roots of the cubic equation $\lambda^3+a\lambda^2+b\lambda+c=0$ will have negative real parts if and only if $a>0$, $c>0$ and $ab>c$. It follows immediately that the necessary and sufficient condition for the eigenvalues to have negative real parts is  $\beta_2>\beta_1$ and  $-\beta_1 f_2 > \beta_2(\theta-f_1)$. Therefore, the endemic equilibrium at the point $C$ is locally asymptotically stable when $f_0>0$, $\beta_2>\beta_1$ and $-\beta_1 f_2 > \beta_2(\theta-f_1)$. The shaded region in Figure 3 shows the $(f_1,\theta)$ pairs for which the endemic equilibrium  is always locally asymptotically stable.

However, if two of the eigenvalues of the Jacobian matrix in (3.7) at the point $C$ are pure 
imaginary complex conjugate and one is negative real, then the linear system at the endemic equilibrium is stable. To examine the conditions for  stability, the eigenvalues  at $C$ are defined as $\lambda^*_1=-a$ and $\lambda^*_{2,3}=\mp bi$ where $a,b>0$. Then, by equating the characteristic equation  $\lambda^{*^3}+a\lambda^{*^2}+b^2\lambda^*+ab^2=0$ to the equation in  (3.9) and using the restrictions on $a$, $b$ and $c$, one finds $-\beta_1 f_2 > \beta_2(\theta-f_1)$ and $\beta_1=\beta_2$.

The conditions for asymptotic stability and stability at the endemic equilibrium in terms of the basic production number are as follows:

$\bullet$ If $f_0>0$, $R_0=R_1$ and $\beta_2>\beta_1$, then the endemic equilibrium is asymptotically stable.

$\bullet$ If $f_0>0$, $R_0=R_1$ and $\beta_2=\beta_1$, then the endemic equilibrium is  stable.

To demonstrate theoretical results for asymptotic stability at the endemic equilibrium,   solution of the system for suitable parameters is given on the left panel in  Figure 4.
The phase portrait corresponding to this solution and its projection curves on  $SI_1$ and $SI_2$ planes are given in Figure 5 and 6, respectively. These figures show that such an equilibrium is asymptotically stable. In addition, phase portraits for the same parameter values but with different initial conditions are given in Figure 9. It is observed that  different initial values do not change the asymptotic stability but they affect the amplitude of solutions.
 
To illustrate the theoretical results for  stability at the endemic equilibrium, solution of the system for suitable parameters is given on the right panel in  Figure 4. 
The phase portrait corresponding to this solution and its projection curves on  $SI_1$ and $SI_2$ planes are given in Figures 7 and 8, respectively. As can be seen in these figures, such an equilibrium is  stable. Additionally,  phase portraits for the same parameter values but with different initial conditions are given in Figure 10. It is observed that different initial values do not change the  stability  but they affect the amplitude 
of the solutions.

\section{Analysis of a Special Case}
 In this section, a special case of the system defined by the relations in (2.1) is considered, and the exact solution of the new system is found.
 
\subsection{Reduced Model} 
 The reduced system considered in this section is obtained when  the system in (2.1) satisfies the following conditions
 
$\bullet$ the transmission rates are identical ($\beta_1=\beta_2$),

$\bullet$
the birth and death rates of the susceptible population, $I_1$-infected group and  $I_2$-infected group   are equal; that is,  $f_0=0$, $f_1=0$ and $f_2=0$.

The corresponding system of nonlinear differential equations is defined by 
\begin{eqnarray}
\begin{aligned}
S'&=-\beta S\,(I_1+I_2)\\
I_1'&=(\beta S-\theta ) \,I_1\\
I_2'&=\beta S\, I_2+\theta I_1
\end{aligned}
\end{eqnarray}
where  $S+I_1+I_2=N_0$. 

\subsection{Exact Solution}
To obtain the exact solution of this reduced case, a new variable
\begin{eqnarray}T'=I_1+I_2
\end{eqnarray}
is defined. If $S$ is considered as a function of $T$, one obtains the following by using the first  equation in (4.1) and the equation in (4.2) together with the initial conditions $S_0$, $I_{1,0}$, $I_{2,0}$ and $T_0$
\begin{eqnarray}
\begin{aligned}
S=S_0\exp(-\beta (T-T_0)).
\end{aligned}
\end{eqnarray}
Substituting (4.3) in the equation $S+I_1+I_2=N_0$ gives
\begin{eqnarray}
\begin{aligned}
I_1+I_2=N_0-S_0\exp(-\beta (T-T_0)).
\end{aligned}
\end{eqnarray}
Substituting of  (4.4) in  (4.2) and then integrating the resulting equation yield
\begin{eqnarray}
\begin{aligned}
T=T_0+\displaystyle \frac {1}{\beta}\ln\bigg[\bigg (1-\displaystyle \frac {S_0}{N_0}\bigg) \exp (\beta N_0 \,t)+\displaystyle \frac {S_0}{N_0}\bigg].
\end{aligned}
\end{eqnarray}
In a similar manner, if $I_1$ is also considered as a function of $T$, one gets the following by using the second equation in (4.1), the  equations in (4.2) and (4.3)
\begin{eqnarray}
\begin{aligned}
\displaystyle (I_1+I_2)\,\frac{dI_1}{dT}=(\beta S_0\exp(-\beta (T-T_0))-\theta)\, I_1.
\end{aligned}
\end{eqnarray}
Substituting of (4.4)  in (4.6) and then integrating the resulting equation yield
\begin{eqnarray}
\begin{aligned}
I_1=I_{1,0} \exp(-\beta (T-T_0)) \bigg( \displaystyle \frac {N_0\,\exp(\beta (T-T_0))-S_0}{N_0-S_0}\bigg)^{ 1-\frac {\theta}{\beta N_0}}.
\end{aligned}
\end{eqnarray}
 If (4.7) is substituted in (4.4), one gets
\begin{eqnarray}
\begin{aligned}
I_2=N_0-S_0 \exp(-\beta (T-T_0))-
I_{1,0} \exp(-\beta (T-T_0)) \bigg( \displaystyle \frac {N_0\,\exp(\beta (T-T_0))-S_0}{N_0-S_0}\bigg)^{ 1-\frac {\theta}{\beta N_0}}.
\end{aligned}
\end{eqnarray}

If (4.5) is replaced in (4.3), (4.7) and (4.8),  the exact solution of (4.1) is expressed as follows
\begin{eqnarray}
\begin{aligned}
S&=\displaystyle \frac {S_0\,\exp{(-\beta N_0 t)}}{1-( {S_0}/{N_0}) +( {S_0}/{N_0})\,\exp{(-\beta N_0 t)}}\\ 
I_1&=\displaystyle \frac {I_{1,0}\,\exp{(-\theta t)}} {1-( {S_0}/{N_0})+( {S_0}/{N_0})\,\exp{(-\beta N_0 t)}}\\ 
I_2&=N_0-\displaystyle \frac {S_0\,\exp{(-\beta N_0 t)}+I_{1,0}\,\exp{(-\theta t)}} {1-( {S_0}/{N_0})+( {S_0}/{N_0})\,\exp{(-\beta N_0 t)}}.\\ 
\end{aligned}
\end{eqnarray}

Graphs for the exact solution are given in Figure 11 for suitable parameters. As can be seen in Figure 11, the susceptible population disappears over time. A period of time after this disappearance, $I_1$-infected group also declines to zero. However, $I_2$-infected group survives.


\section{Conclusion}
In this paper, a mathematical study describing a new susceptible-infected model is  presented. The $SI_1I_2$ epidemic model has two different infectious groups  which have different clinical forms of  infection.  Members of one of the infectious group  may become a member of the other infectious group during the progress of the illness. It is assumed that the illness has no cure and therefore individuals who are infected will eventually die of the disease or some other unrelated cause.
\par
Initially, positiveness of the solutions of the $SI_1I_2$ system is proved. In addition, the basic reproduction number  which has an important role in the epidemiology of a disease, and  the   equilibrium points are found.  It is shown that the system may have  three  equilibrium points.  Furthermore, the stability conditions of the equilibrium points are obtained in terms of the parameters. The phase portraits for asymptotic stability and stability at the  endemic equilibrium are given for suitable parameters and with different initial conditions.

Stability analysis  of the  equilibrium points reveals the following aspects:

 (1) The species  will become extinct if the birth rate is smaller than the death rate in the susceptible population in the presence of  infection.
 
 (2) One form of the infection  ($I_2$) may persist  while the other form ($I_1$) dies out if the birth rate is greater than the death rate in the susceptible population, and if the basic reproduction number of the system is equal to the basic reproduction  number of the group $I_2$.
 
 (3) Endemic equilibrium may exist if the  birth rate is greater than the death rate in the susceptible population, and if the contact rate of $I_2$ is greater than the contact rate of $I_1$, and if the basic reproduction number of the system is equal to the basic reproduction  number of the group $I_1$.
 
A special case for specific parameters is investigated.  The exact solution of this reduced system is also obtained. Examination of this system reveals that $I_2$-infected group survives while other groups disappear over time.

\section*{References}

\begin{figure}[ht]
\centering
\includegraphics[scale=0.5]{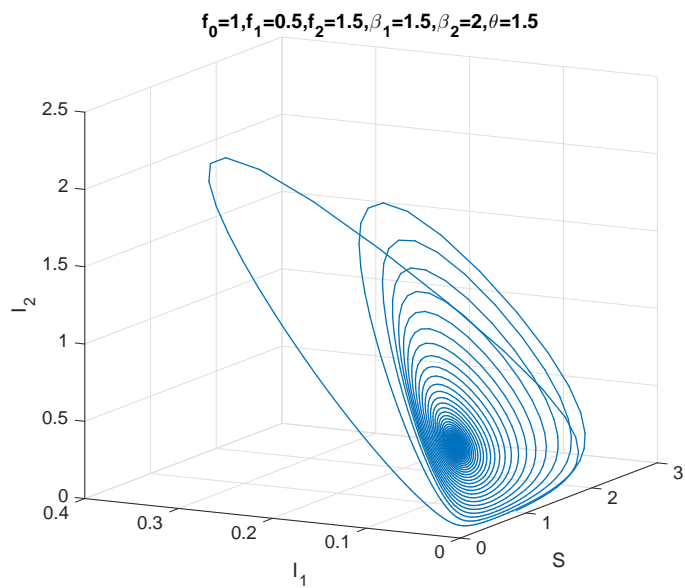}
\caption{Phase portrait for asymptotic stability at the endemic equilibrium on $SI_1I_2$ space.}
\end{figure}

\begin{figure}[ht]
\centering
\includegraphics[scale=0.5]{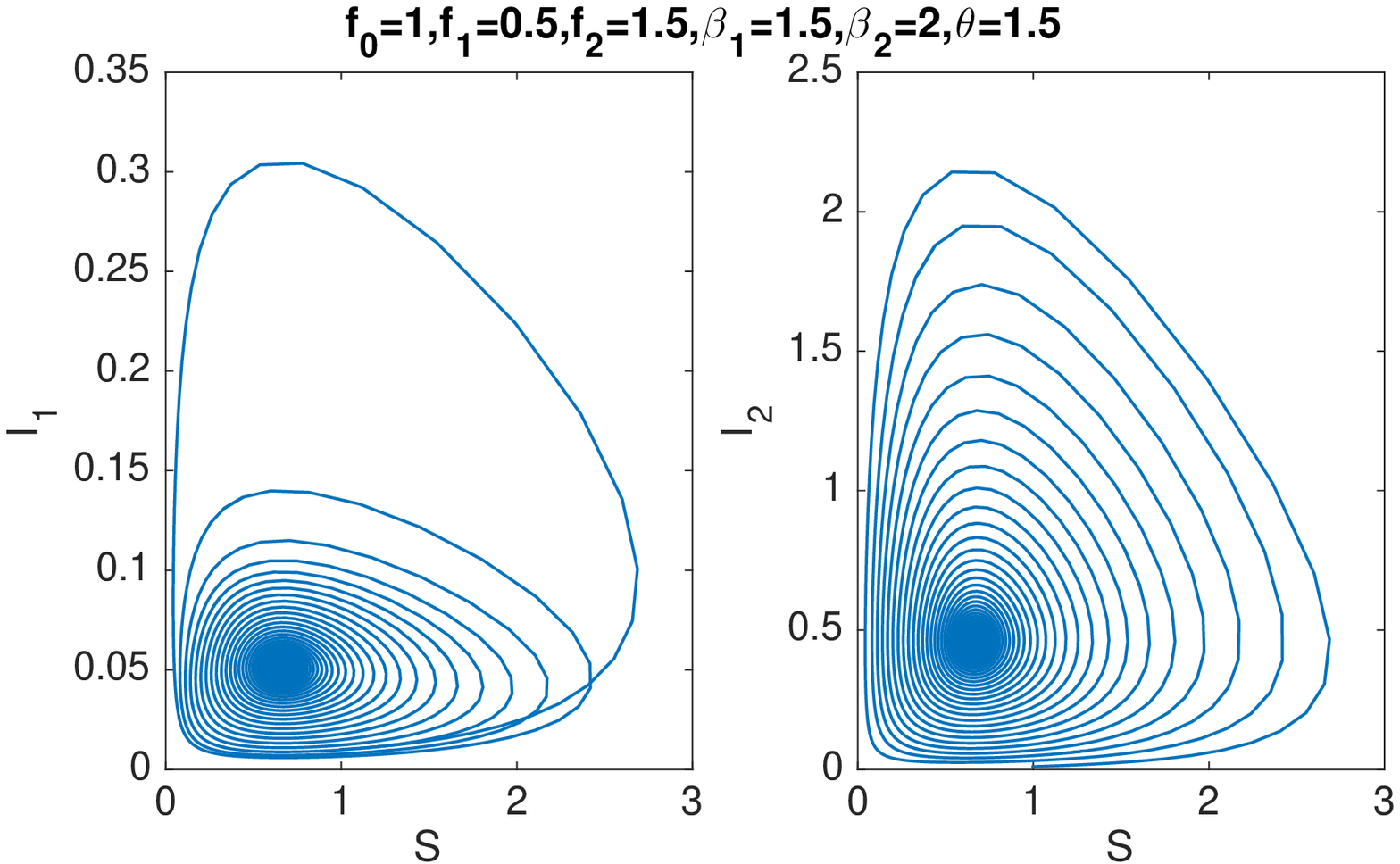}
\caption{Projection  of the phase portrait for asymptotic stability at the endemic equilibrium on $SI_1$ and $SI_2$ plane.}
\end{figure}

\begin{figure}[ht]
\centering
\includegraphics[scale=0.5]{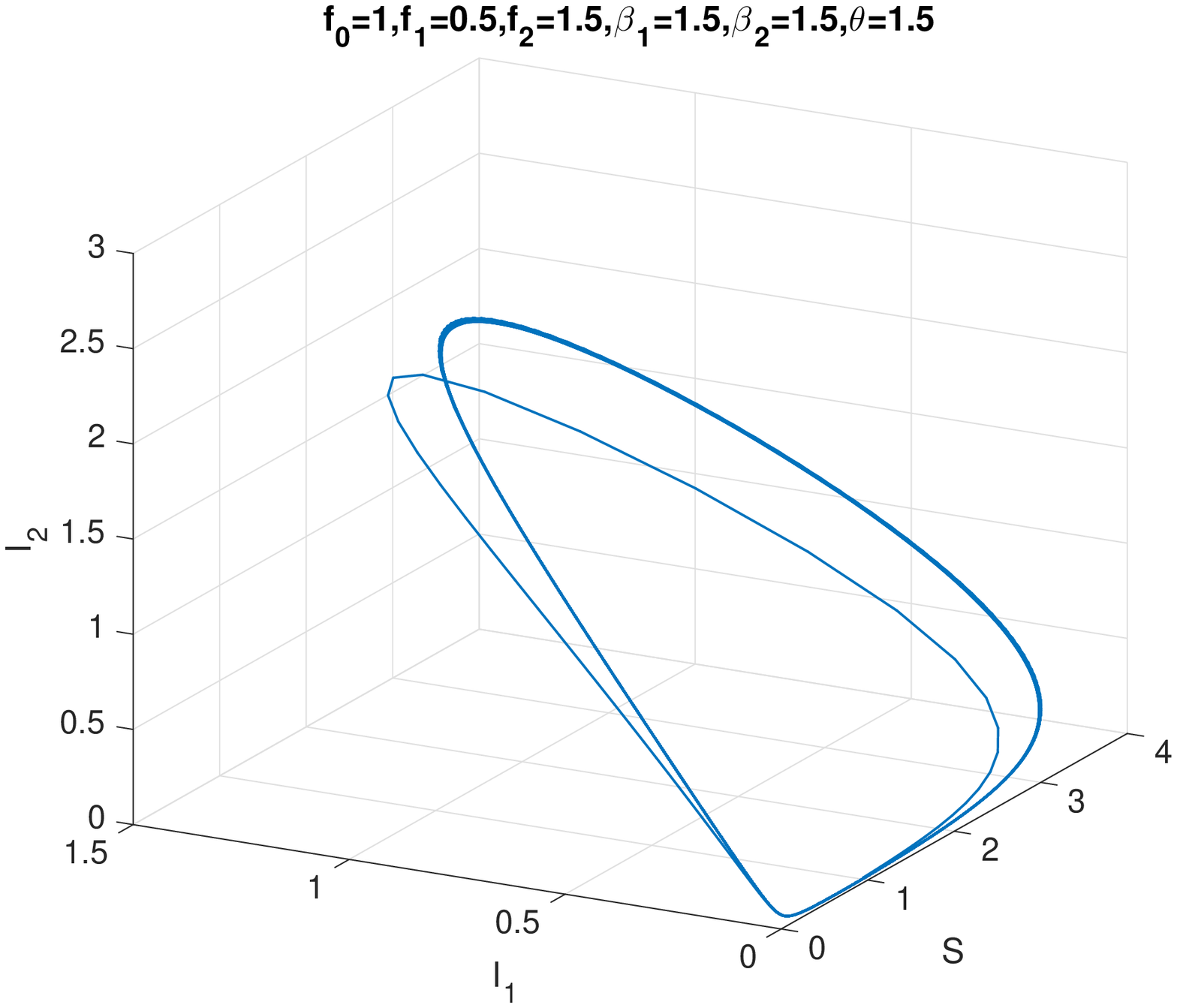}
\caption{Phase portrait for stability at the endemic equilibrium on $SI_1I_2$ space.}
\end{figure}

\begin{figure}[ht]
\centering
\includegraphics[scale=0.5]{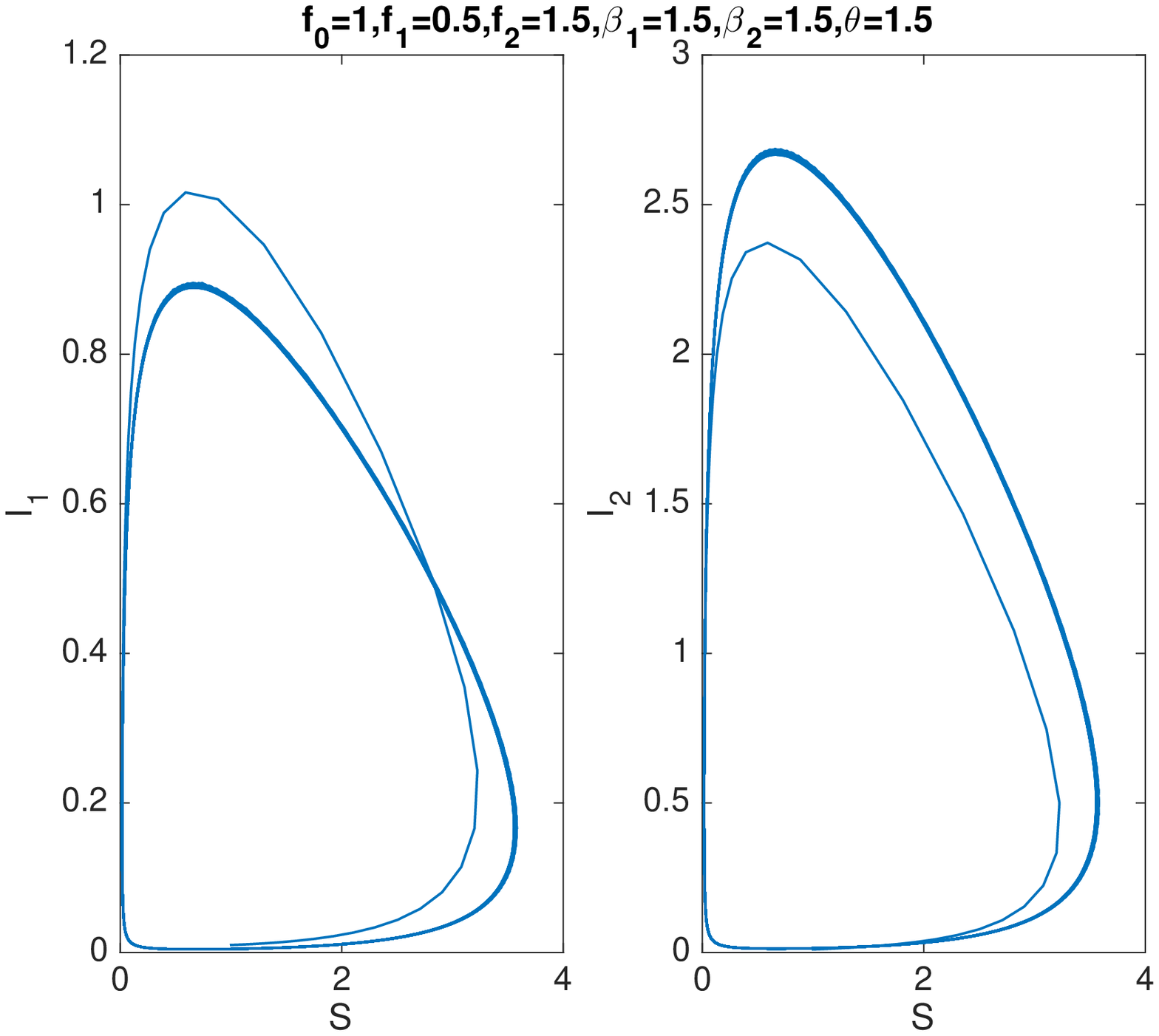}
\caption{Projection of the phase portrait for  stability at the endemic equilibrium on $SI_1$ and $SI_2$ plane.}
\end{figure}

\begin{figure}[ht]
\centering
\includegraphics[scale=0.5]{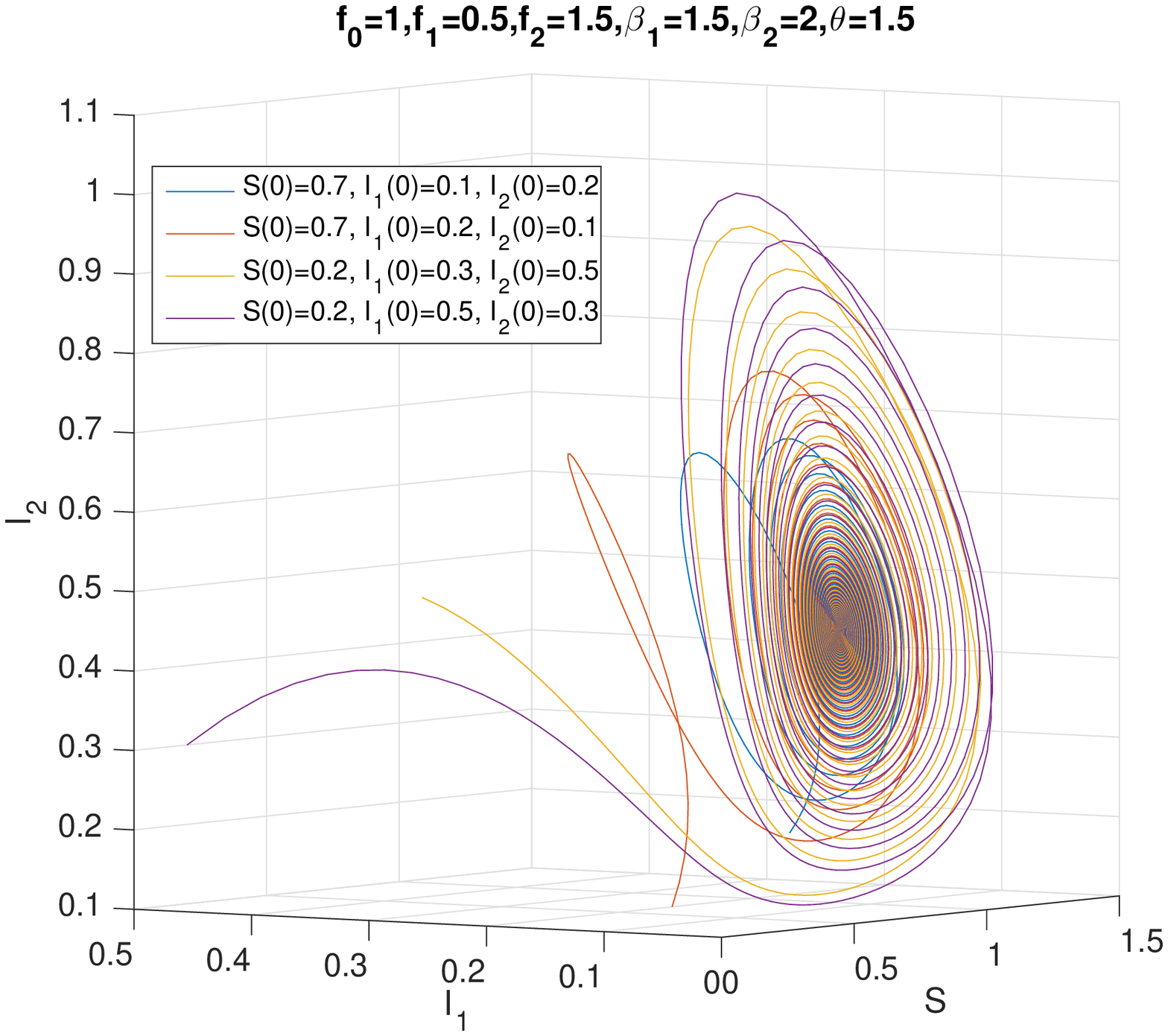}
\caption{Phase portrait for asymptotic stability at the endemic equilibrium for different initial conditions.}
\end{figure}

\begin{figure}[ht]
\centering
\includegraphics[scale=0.5]{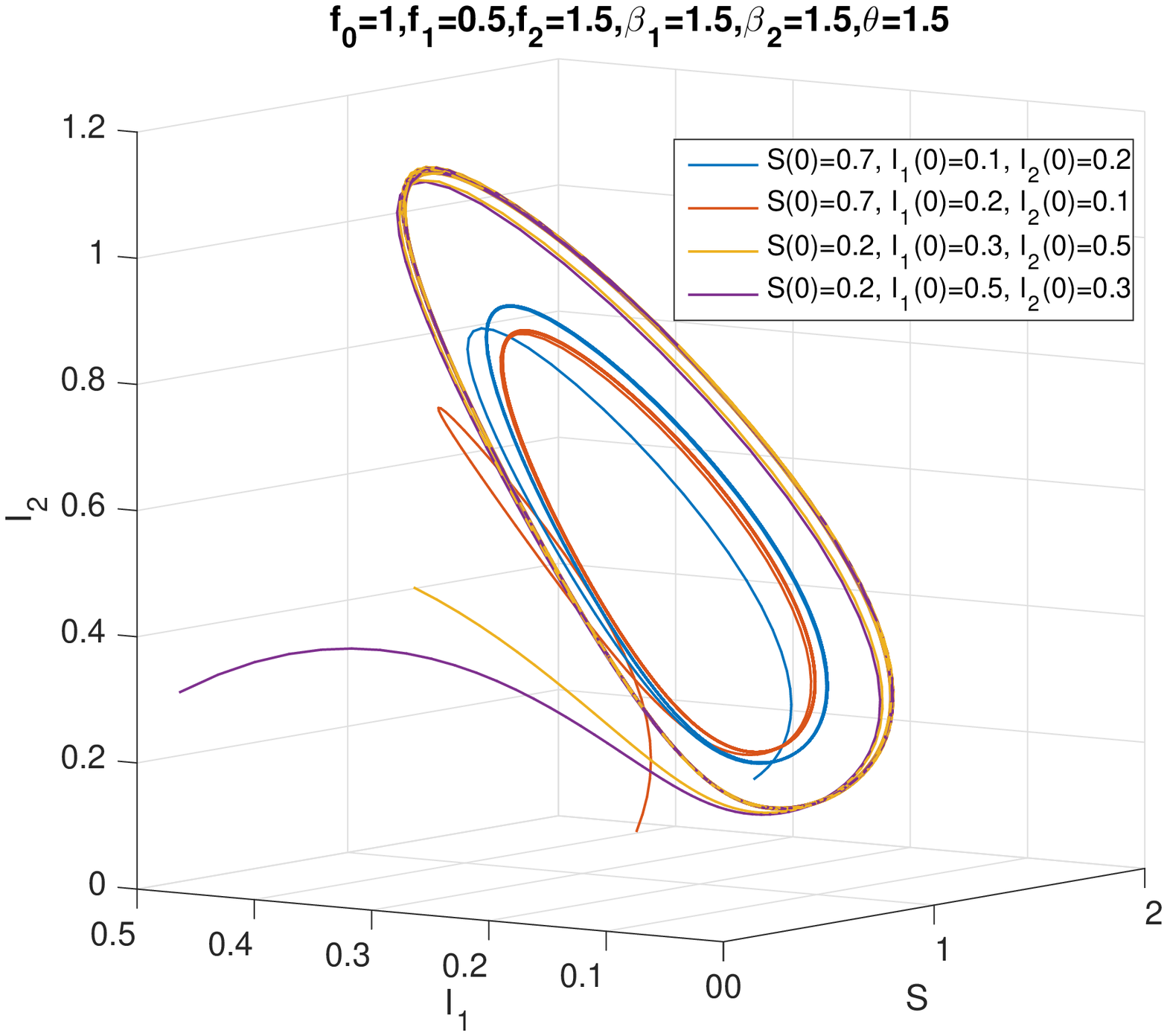}
\caption{Phase portrait for stability at the endemic equilibrium for different initial conditions.}
\end{figure}

\begin{figure}[ht]
\centering
\includegraphics[scale=0.5]{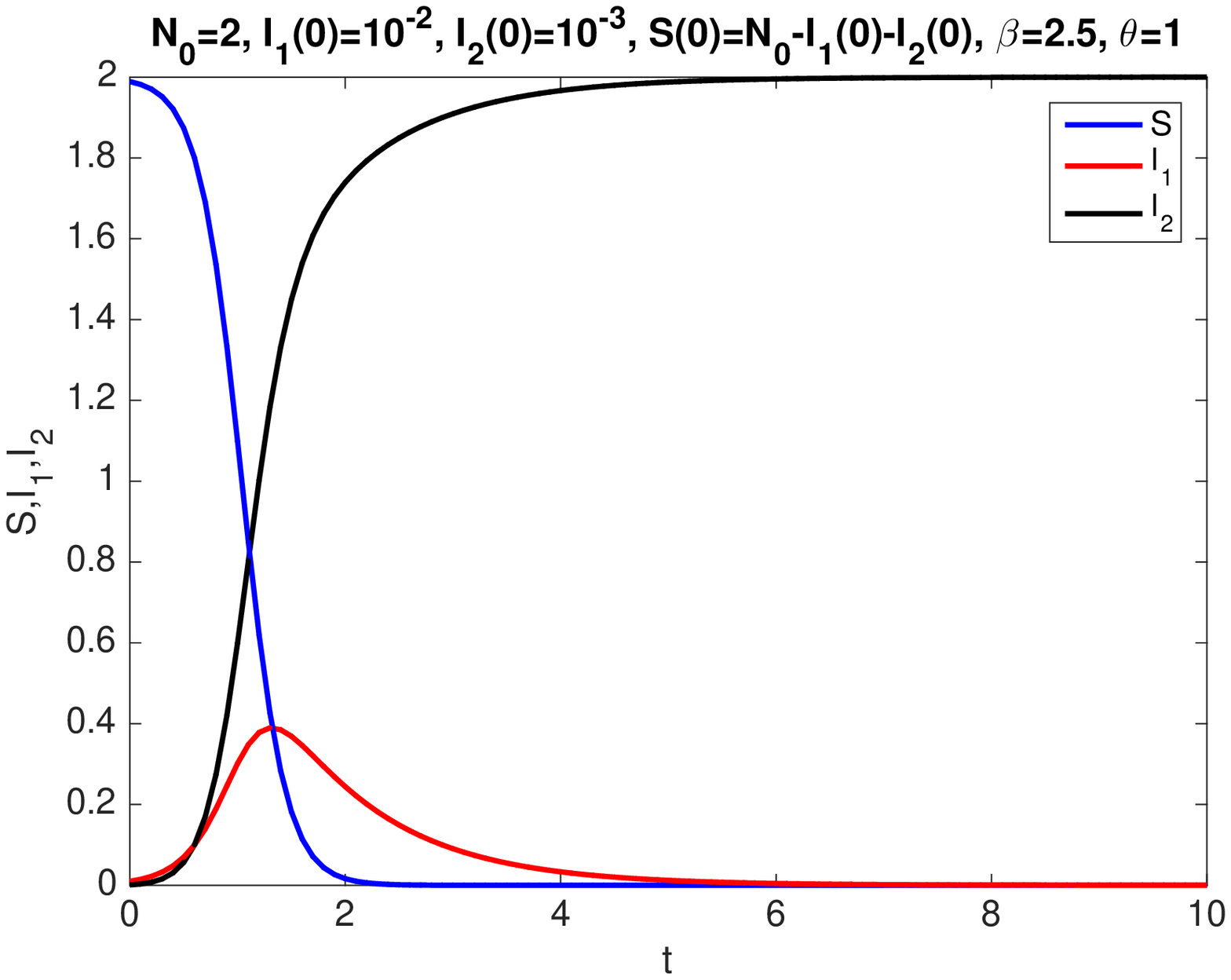}
\caption{Graphs of $S$, $I_1$ and $I_2$ for the exact solution of the reduced model for suitable parameters.}
\end{figure}


\begin{thebibliography}{9}

\bibitem{ker}
Kermack, WO,  McKendrick AG. A contribution to the mathematical theory of epidemics. \textit{P R Soc ond A-Conta}. 1927; 115(772):700-721. DOI: 10.1098/rspa.1927.0118.

\bibitem{cia} Ciarcia C, Falsaperla P, Giacobbe A, Mulone G. A mathematical model of anorexia and bulimia. \textit{Math Method Appl Sci}. 2015; 38(14): 2937-2952. DOI: 10.1002/mma.3270.

\bibitem{whi} White E, Comiskey C. Heroin epidemics, treatment and ODE modelling.\textit{Math Biosci.} 2007; 208(1): 312-324. DOI:10.1016/j.mbs.2006.10.008.

\bibitem{cho} Chowell G, Fenimore PW, Castillo-Garsow MA, Castillo-Chavez C. SARS outbreaks in Ontario, Hong Kong and Singapore: the role of diagnosis and isolation as a control mechanism.\textit{J Theor Biol.} 2003; 224(1): 1-8. DOI: 10.1016/S0022-5193(03)00228-5.

\bibitem{gum} Gumel AB, Ruan S, Day T, Watmough J, Brauer F, Van den Driessche P, Gabrielson D,  Bowman C, Alexander ME, Ardal S, Wu J, Sahai BM. Modelling strategies for controlling SARS outbreaks. \textit{P Roy Soc Lond B Bio.} 2004; 271(1554): 2223-2232. DOI: 10.1098/rspb.2004.2800.

\bibitem{bau} Brauer F. The Kermack-McKendrick epidemic model revisited.\textit{Math Biosci.} 2005; 198(2): 119-131. DOI: 10.1016/j.mbs.2005.07.006.

\bibitem{rao} Rao DM, Chernyakhovsky A, Rao V. Modeling and analysis of global epidemiology of avian influenza.\textit{Environ Modell Softw.} 2009; 24(1): 124-134. DOI: 10.1016/j.envsoft.2008.06.011.

\bibitem{yan} Yan-li LI. Study on SI Transmission Model of Highly Pathogenic Avian Influenza. \textit{Journal of Anhui Agricultural Sciences.} 2009; 28.

\bibitem{che} Che S, Xue Y, Ma L. The stability of highly pathogenic avian influenza epidemic model with saturated contact rate. \textit{Appl Math.} 204; 5(21): 3365. DOI:10.4236/am.2014.521313.

\bibitem{pek} Dobie AP, Demirci A, Bilge AH, Ahmetolan S. On the time shift phenomena in epidemic models. \textit{arXiv.} 2019;  arXiv:1909.11317.

\bibitem{dem} Demirci A, Dobie AP, Bilge AH, Ahmetolan S. Unexpected parameter ranges of the 2009 A (H1N1) epidemic for Istanbul and the Netherlands.\textit{ arXiv.} 2020; arXiv:2001.10351.

\bibitem{cob} Coburn BJ, Wagner BG, Blower S. Modeling influenza epidemics and pandemics: insights into the future of swine flu (H1N1).\textit{ BMC Medicine.} 2009; 7(1): 30. DOI: 10.1186/1741-7015-7-30. 

\bibitem{chr} Chretien JP, Riley S, George DB. Mathematical modeling of the West Africa Ebola epidemic.\textit{ Elife.} 2015; 4: e09186. DOI: 10.7554/eLife.09186.

\bibitem{webb} Webb G, Browne C, Huo X, Seydi O, Seydi M, Magal P. A model of the 2014 Ebola epidemic in West Africa with contact tracing.\textit {PLoS Curr.} 2015; 7. DOI: 10.1371/currents.outbreaks.846b2a31ef37018b7d1126a9c8adf22a.

\bibitem{mam} Mamo DK, Koya PR. Mathematical modeling and simulation study of SEIR disease and data fitting of Ebola epidemic spreading in West Africa.\textit{ JMEST}. 2015; 2(1): 106-114.

\bibitem{ahm} Ahmetolan S, Bilge AH, Demirci A, Dobie AP, Ergonul O. What Can We Estimate from Fatality and Infectious Case Data? A case Study of Covid-19 Pandemic.\textit{ arXiv.} 2020; arXiv:2004.13178.

\bibitem{kuc} Kucharski AJ, Russell TW, Diamond C, Liu Y, Edmunds J, Funk S, Eggo RM. Early dynamics of transmission and control of COVID-19: a mathematical modelling study. \textit{Lancet Infect Dis}. 2020; 20(5): 553-558. DOI: 10.1016/S1473-3099(20)30144-4.

\bibitem{liu} Liu  Z, Magal P, Seydi O, Webb G. A COVID-19 epidemic model with latency period. \textit{Infec Dis Mod}. 2020; 5: 323-337. DOI: 10.1016/j.idm.2020.03.003.

\bibitem{and}
Anderson RM, Anderson B, May RM. \textit{Infectious diseases of humans: dynamics and control}. UK: Oxford University Press; 1992.

\bibitem{mur}
Murray JD. \textit{Mathematical biology: I. An introduction}. USA: Springer; 2007.

\bibitem{Van}
Van den Driessche P, Watmough J. Reproduction numbers and sub-threshold endemic equilibria for compartmental models of disease transmission. \textit{Math Biosci.} 2002; 180(1-2): 29-48. DOI: 10.1016/ S0025-5564(02)00108-6. 

\bibitem{Die}
Diekmann O, Heesterbeek JAP, Metz JAJ. On the definition and the computation of the basic reproduction ratio R 0 in models for infectious diseases in heterogeneous populations. \textit{J Math Bio.} 1990; 28(4): 365-382. DOI: 10.1007/BF00178324.





\end{thebibliography}
\end{document}